# Observation of Van Hove singularities in twisted graphene layers

Guohong Li[1], A. Luican[1], J.M. B. Lopes dos Santos[2], A. H. Castro Neto[3], A. Reina[4], J. Kong[5] and E.Y. Andrei[1]

[1] Department of Physics and Astronomy, Rutgers University, Piscataway, NJ 08855

[2] CFP and Departamento de Física, Faculdade de Ciências Universidade do Porto, 4169-007 Porto, Portugal

[3] Department of Physics, Boston University, 590 Commonwealth Avenue, Boston, MA 02215

[4] Department of Materials Science and Engineering, MIT, Cambridge, MA 02139

[5] Department of Electrical Engineering and Computer Science, MIT, Cambridge, MA 02139

Electronic instabilities at the crossing of the Fermi energy with a Van Hove singularity[1] in the density of states often lead to new phases of matter such as superconductivity[2,3], magnetism[4] or density waves[5]. However, in most materials this condition is difficult to control. In the case of single-layer graphene, the singularity is too far from the Fermi energy[6] and hence difficult to reach with standard doping and gating techniques[7]. Here we report the observation of low-energy Van Hove singularities in twisted graphene layers seen as two pronounced peaks in the density of states measured by scanning tunneling spectroscopy. We demonstrate that a rotation between stacked graphene layers can generate Van Hove singularities, which can be brought arbitrarily close to the Fermi energy by varying the angle of rotation. This opens intriguing prospects for Van Hove singularity engineering of electronic phases.

In two dimensions, a saddle point in the electronic band structure leads to a divergence in the density of states, also known as a Van Hove singularity[1] (VHS). When the Fermi energy ($E_F$) is close to the VHS, interactions, however weak, are magnified by the enhanced density of states (DOS), resulting in instabilities, which can give rise to new phases of matter[2, 3, 4, 5] with desirable properties. This implies the possibility of engineering material properties by bringing $E_F$ and the VHS together. However, in most materials one cannot change the position of the VHS in the band structure. Instead, it may be possible to tune $E_F$ through the VHS by chemical doping[8] or by gating[7]. In this regard, graphene, the recently discovered two-dimensional form of carbon, is quite special[5, 9]. It has linearly dispersing bands at the K (K') point in the Brillouin zone, the so-called Dirac points, and a DOS that is linear and vanishes at Dirac point. The fact that this material is truly two-dimensional and has a low DOS means that it cannot screen applied electric fields, allowing for strong gating and ambipolar behaviour[7]. However, although the band structure of graphene[5] contains a VHS, its large distance from the Dirac point makes it prohibitively difficult to reach by either gating or chemical doping. We show that by introducing a rotation between stacked graphene layers, it is possible to induce VHSs that are within the range of $E_F$ achievable by gate tuning. As the samples studied here are not intentionally doped, $E_F$ is within a few millielectronvolts of the Dirac point.

Rotation between graphene layers is often observed as a Moiré pattern on graphite surfaces[10], as illustrated in Fig. 1. Graphite consists of stacked layers of graphene, the lattice structure of which contains two interpenetrating triangular sublattices, denoted A and B. In the most common (Bernal) stacking, adjacent layers are shifted by one atomic spacing, so that B atoms of layer 2 (B2) sit directly on top of A atoms of layer 1 (A1) and B1 and A2 atoms are in the center of the hexagons of the opposing layer. Assuming a rotation through an angle $\theta$ about an A1 (B2) site in bilayer graphene, a set of conditions for commensurate periodic structures leading to Moiré patterns is easily derived[11] $\cos(\theta_i) = (3i^2 + 3i + 1/2)/(3i^2 + 3i + 1)$, with $i$ an integer ($i=0$, $\theta = 60°$ corresponds AA stacking and $i \to \infty$, $\theta = 0°$ to AB stacking) and lattice constant of the superlattice $L = a_0\sqrt{3i^2 + 3i + 1}$ where $a_0 \sim 2.46$Å is the atomic lattice constant. For the extended Moiré pattern in Fig.1, $\theta_i$=1.79°, is obtained from the super-lattice period, $L$=7.7±0.3nm, corresponding to $i$=18.

Thus far, most work on Moiré patterns focused on structural aspects revealed by scanning tunneling microscopy (STM), but their effect on the electronic properties has not been addressed. By using scanning tunneling spectroscopy (STS), we find that rotation markedly alters the DOS. Figure 2 shows the tunneling differential conductance, d$I$/d$V$, a quantity proportional to the local DOS (ref. 12). In regions inside the Moiré pattern (M1 and M2), the spectra develop two sharp peaks flanking the Dirac points with energy separation $\Delta E_{vhs} \sim 82 meV$. Below, we show that these peaks correspond to rotation-induced VHSs. The effect of the VHS on the DOS is evident in the energy dependence of the d$I$/d$V$ maps. Close to the VHS, the maps develop a strong density modulation (Figs 2f and 3c), characteristic of charge-density waves (CDWs), suggesting an impending Fermi-surface instability. In contrast, for energies away from the VHS, the charge density becomes nearly homogeneous.

To explore the angle dependence of $\Delta E_{vhs}$, we studied graphene layers prepared by chemical vapour deposition[13] (CVD). CVD graphene layers have a strong twisting tendency revealed by Moiré patterns with a range of rotation angles. The example in Fig. 3a shows a pattern, corresponding to $\theta \sim 1.16°$. The STS spectra in this region reveal strong VHSs (Fig. 3b) with a much smaller $\Delta E_{vhs} \sim 12 meV$. The pronounced spatial modulation of the d$I$/d$V$ maps at energies close to these VHSs (Fig. 3c), indicating the formation of a CDW, is significantly stronger than for the pattern in Fig. 2f, where the more widely separated VHSs are farther away from the Fermi energy. For a pattern with an even larger angle, ~3.4°, and $\Delta E_{vhs} \sim 430 meV$, the localization is weaker still (see Supplementary Fig. S3), consistent with theoretical predictions[14].

We next show that the VHSs are induced by the rotation and use the model developed in ref. 11 to derive the angle dependence of $\Delta E_{vhs}$. A rotation between two graphene layers causes a shift between the corresponding Dirac points in momentum space, so that the Dirac wavevector of the rotated layer is $K^\theta = K + \Delta K$, where $\Delta K = K \times 2\sin(\theta/2)$. If we use the same origin of momentum for the two layers, so that a uniform hopping couples states of the same momentum in both layers, the zero-energy states do not occur at $k=0$, but rather at $k = -\Delta K/2$ in layer 1 and $k = \Delta K/2$ in layer 2. Unlike in the AB stacked bilayer, there is no direct coupling of the zero-energy states of one layer to the zero-energy states of the other. As shown in ref. 11, the states near the Dirac cone of each layer couple with amplitudes of order $t_\perp^\theta \approx 0.4 t_\perp$ to states of energy $\pm \hbar v_F \Delta K$ of the opposing layer and the linear dispersion is preserved near zero energy. Here, $t_\perp$ is the interlayer hopping for un-rotated layers and $v_F \sim 10^6$ m/s the Fermi velocity. The two Dirac cones intersect near the center of the superlattice Brillouin zone and hybridize, resulting in a saddle point in the energy dispersion and in two symmetric VHSs (Fig. 4a).

This mechanism of rotation-induced low-energy VHSs is not restricted to the bilayer. We extended the calculation to a trilayer consisting of an AB stacked bilayer and a third rotated layer on top and again found two VHSs in the DOS (Fig. 4b). In both cases $\Delta E_{vhs}$ is controlled by the energy scale $\hbar v_F \Delta K \propto \theta$. Figure 4c shows that this model accounts well for the experimental angle dependence of $\Delta E_{vhs}$.

This analysis shows that the mechanism behind the formation of VHSs is quite robust, and we expect it to apply more generally whenever one layer is rotated with respect to the others, even for a graphene flake over graphite. The strong twisting-angle dependence of $\Delta E_{vhs}$ is its unmistakable signature.

In the bilayer, the two Van Hove peaks are exactly symmetrical, whereas in the trilayer, a slight asymmetry occurs because of the third layer, but not as large as that observed experimentally. The discrepancy can be attributed to two factors. First, the experiments probe the local DOS, which varies across the unit cell of the Moiré pattern, whereas the calculation refers to the total DOS. Averaging the data over several unit cells reduces the asymmetry. Second, a bias between the layers can also enhance the asymmetry. In the AB stacked bilayer, a perpendicular electric field opens a gap in the spectrum[15]; not so in the twisted bilayer. As states close to zero energy result from the hybridization of zero-energy states in one layer with states of energy $\pm \hbar v_F \Delta K$ in the other, the corresponding wavefunctions have different weights in the two layers; when an electric field is applied, the two Dirac points move in opposite directions in energy[11]. The positive and negative energy states at the saddle points now have different weights in the two layers. As the STM probes predominantly the top layer, we expect this asymmetry in the STS. This effect of the bias in the DOS of the bilayer is shown in Fig. 4b (mid-panel).

It is important to note that the VHSs can form only in the presence of finite interlayer coupling. For vanishingly small interlayer coupling, $t_\perp \sim 0$, as is the case when a detached graphene flake is found on a graphite substrate[16] (see also Supplementary Information), the VHSs will not form even though a Moiré pattern may be visible in the STM images[17] (Supplementary Information). It is well known that on the surface of graphite one often finds graphene flakes loosely bound to the surface where interlayer coupling is completely suppressed[16]. Such flakes, regardless of the presence or absence of rotation with respect to the underlying layer, show an uninterrupted honeycomb structure and a Landau-level sequence characteristic of massless Dirac fermions[16]. However, they do not show VHSs in their zero-field DOS.

The situation here is reminiscent of transition-metal dichalcogenides[14], which form triangular lattices where both CDWs and superconductivity are observed[18]. In fact, these materials have large saddle regions in their band structure[19], conjectured to be the origin of the CDW instability[5]. However, because the saddle regions and the position of $E_F$ are determined by the chemical composition, it is difficult to control or predict where instabilities occur. In contrast, in twisted graphene layers, both the position of $E_F$ and that of the VHSs can be controlled by gating and rotation respectively, providing a powerful toolkit for manipulating electronic phases. To achieve control of $E_F$, the graphene sample can be deposited on a non-conducting substrate, or better yet suspended above a gate[20, 21]. Although our present experimental set-up did not allow tuning of $E_F$, it was possible to find rotated layers in which the VHSs and $E_F$ were sufficiently close to induce a strong CDW. A method to vary $\theta$ *in situ* (and thus to control the position of the VHSs) using an AFM was demonstrated in the discovery of superlubricity in graphite[22]. Superlubricity, which occurs in twisted graphene layers at small rotation angles, was attributed to decoupling of the layers. This led to the assumption that a small twist between layers would decouple them and reveal the electronic structure of an isolated graphene layer[17]. However, counterintuitively, we found that a rotation, however small, markedly affects the band structure and leads to the appearance of VHSs.

The experiments described here demonstrate that in graphene, unlike in any other known material, one can tune the position of the VHSs by controlling the relative angle between layers. As the VHS approaches the Fermi energy, we find a strongly localized CDW corresponding to the wavevector separation between the VHSs. This opens exciting opportunities for inducing and exploring correlated electronic phases in graphene.

**Acknowledgments**. E.Y.A. acknowledges D.O.E. support under DE-FG02-99ER45742. E.Y.A. and A.L. acknowledge support from the Lucent-Alcatel foundation and partial NSF support under NSF-DMR-0906711. A.H.C.N. acknowledges partial support of the US Department of Energy under grant DE-FG02-08ER46512. J.M.B.L.S. acknowledges support from FCT under grant PTDC/FIS/64404/2006. A.R. and J.K. acknowledge partial support of NSF DMR 0845358 and the Materials, Structure and Devices centre. A.H.C.N. and J.K. acknowledge support under ONR MURI N00014-09-1-1063. We thank R. Nahir, K. Novoselov and A. Geim for fabricating the CVD graphene membranes.

**Figure Captions**

**FIG. 1 Scanning tunneling microscopy (STM) of a graphene flake on a freshly cleaved surface of highly oriented pyrolitic graphite (HOPG) revealing a Moiré pattern.**

**a**, Large-area scan of a graphene flake. Regions M2 and G flank the boundary of the Moiré pattern. **b**, Zoom-in of the frame in **a**, showing a Moiré pattern with period 7.7±0.3 nm. Inset: Fourier transforms of the superstructure. **c**, Zoom-in to the center of the pattern, region M1. **d**, Atomic-resolution image of a bright spot. Inset: Fourier transforms of the atomic-resolution image. **e**,**f**, Atomic-resolution image on bright and dark regions of the pattern, showing a well-ordered triangular lattice within the bright spots (**e**) and a less-ordered honeycomb-like structure in between (**f**). The former indicates Bernal-stacked layers[16, 23], whereas the latter suggests slipped stacking, resulting from a small-angle rotation between layers. Tunneling current 20 pA, sample bias voltage 300 mV.

**FIG. 2 Scanning tunneling spectroscopy inside and outside Moiré pattern with $\theta$ =1.79°.**
**a**, Topography near the boundary of the pattern. **b**, Tunneling spectra in the center of the pattern (M1 in Fig. 1c) and at its edge (M2), showing two sharp peaks flanking the Dirac point with $\Delta E_{vhs}$ ~ 82 meV. The peaks are absent outside the pattern area (G), where the spectrum is typical of graphite[24]. **c**, Spatial dependence of tunneling spectra along a line connecting point M2 inside the pattern to point G outside it at positions marked by the white dots in **a**, showing monotonic evolution between the rotated and unrotated regions. Tunneling current 22 pA, sample bias voltage 300 mV, a.c. bias modulation 5 mV$_{rms}$ at 340 Hz. **e**, Comparison of tunneling spectra on bright and dark regions within the pattern at positions indicated in **d**. Although all spectra show two peaks, the peak heights and degree of asymmetry depend on their position in the pattern. **f**, d$I$/d$V$ maps of the region shown in **d**. The top map corresponding to the VHS just below the Fermi energy (-58 meV) shows a strong density modulation, suggesting an imminent Fermi-surface instability. In contrast, the bottom map taken farther from the VHS (77 meV) does not show the modulation.

**FIG. 3 Moiré pattern with $\theta$ =1.16° on CVD graphene**. **a**, Topography taken at 200 mV and 20 pA, showing a pattern with a superlattice constant of 12±0.5 nm, corresponding to $\theta \sim 1.16°$. Scale bar: 2 nm. **b**, Scanning tunneling spectra taken in bright and dark regions of the pattern. a.c. modulation 1 mV$_{rms}$ at 340 Hz. **c–e**, d$I$/d$V$ maps of the region shown in **a** at three bias voltages: far from the VHS (90 mV) (**c**), the map is smooth, indicating extended states; and maps at the energies of the VHS, 1 mV (**d**) and -11 mV (**e**), show localized states. Scale bar: 2 nm

**FIG. 4 Energy separation between Van Hove singularities.** (a) Dispersion of lowest energy states for $\theta$ = 1.79°, $t_\perp \approx 0.27$ eV, in a bilayer. A saddle point (marked "sp") is visible between the two Dirac points in the negative energy band; another one exists in the positive energy band. (b) Top: Density of states of bilayer with VHS corresponding the energy of the saddle points for $\theta$ = 1.79°, $t_\perp \approx 0.24$ eV and no interlayer bias; middle: same with interlayer bias V$_{bias}$ = 0.15 V ; bottom: same for a trilayer and no bias. Insets represent the band energies along a line joining the two Dirac points for the corresponding layer configurations. (c) Angle dependence of $\Delta E_{vhs}$. Open circles: bilayer with $t_\perp$ = 0.27 eV and V$_{bias}$ = 0 V; filled diamonds: trilayer $t_\perp$ = 0.24 eV, V$_{bias}$ = 0 V ; crosses: experimental data. The error in peak separation was estimated from the peak width at 95% of peak height . The error bars are smaller than the symbols at low angles. Both the bilayer and trilayer fit the results, albeit with slightly different values of $t_\perp$. For angles, 2°< $\theta$ <5°, the peak separation is roughly described by $\Delta E_{vhs} = \hbar v_F \Delta K - 2t^\theta$ , but curves upward at smaller angles.

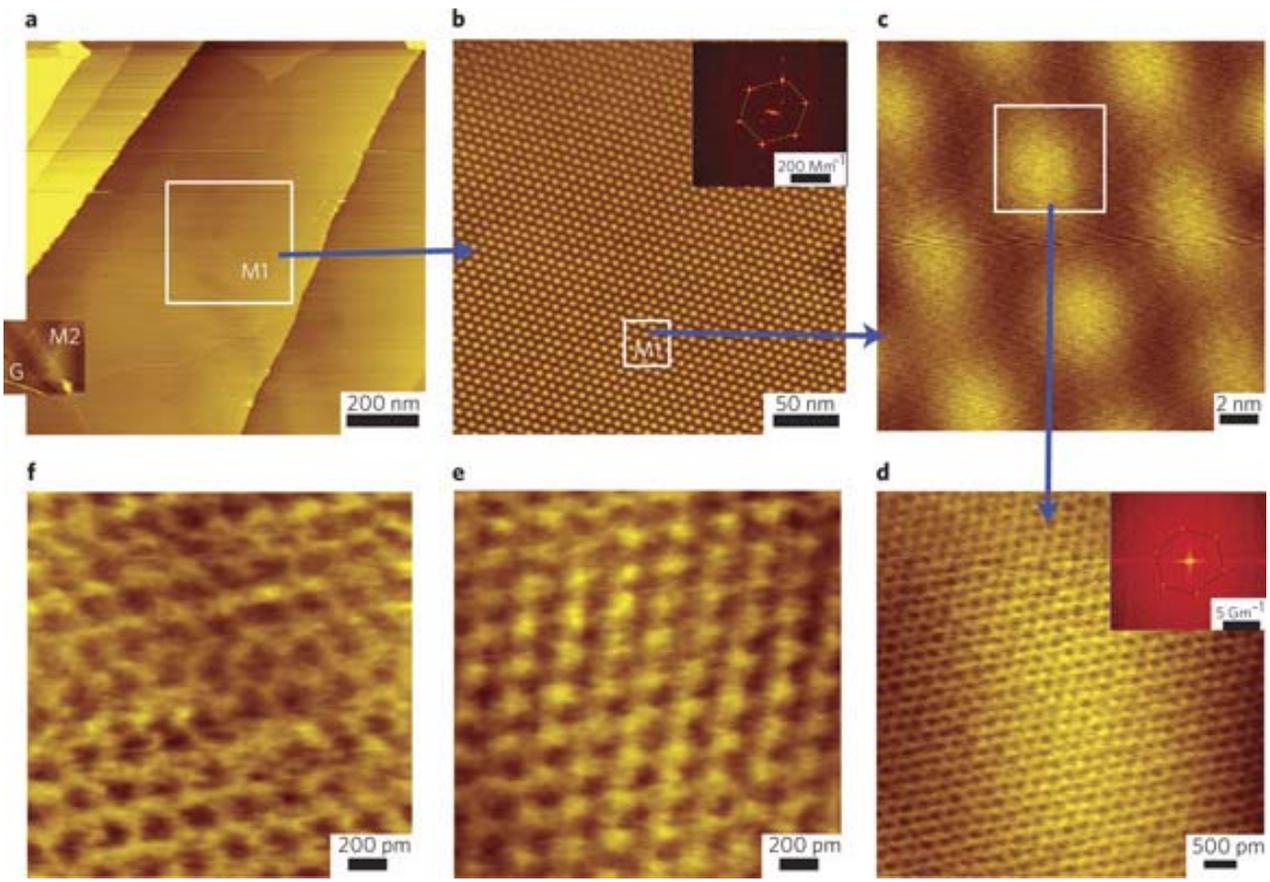

Figure 2

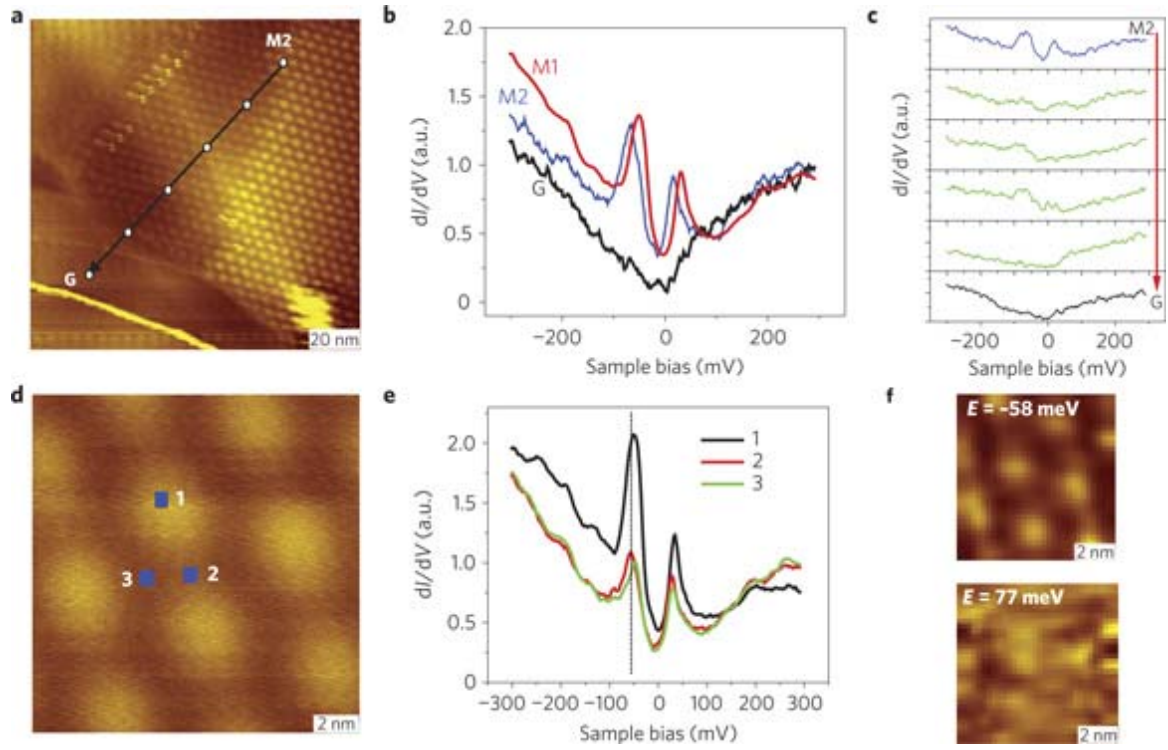

Figure 3

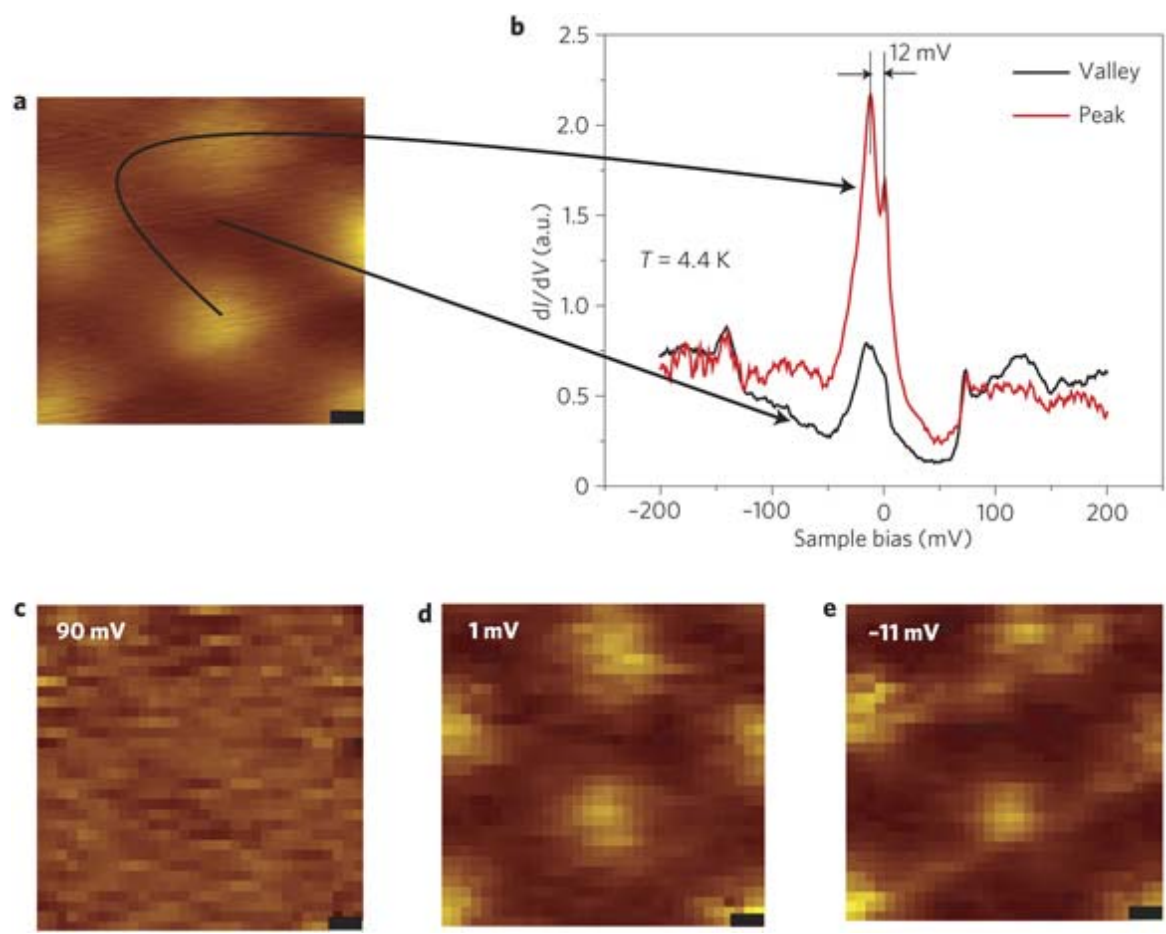

Figure 4

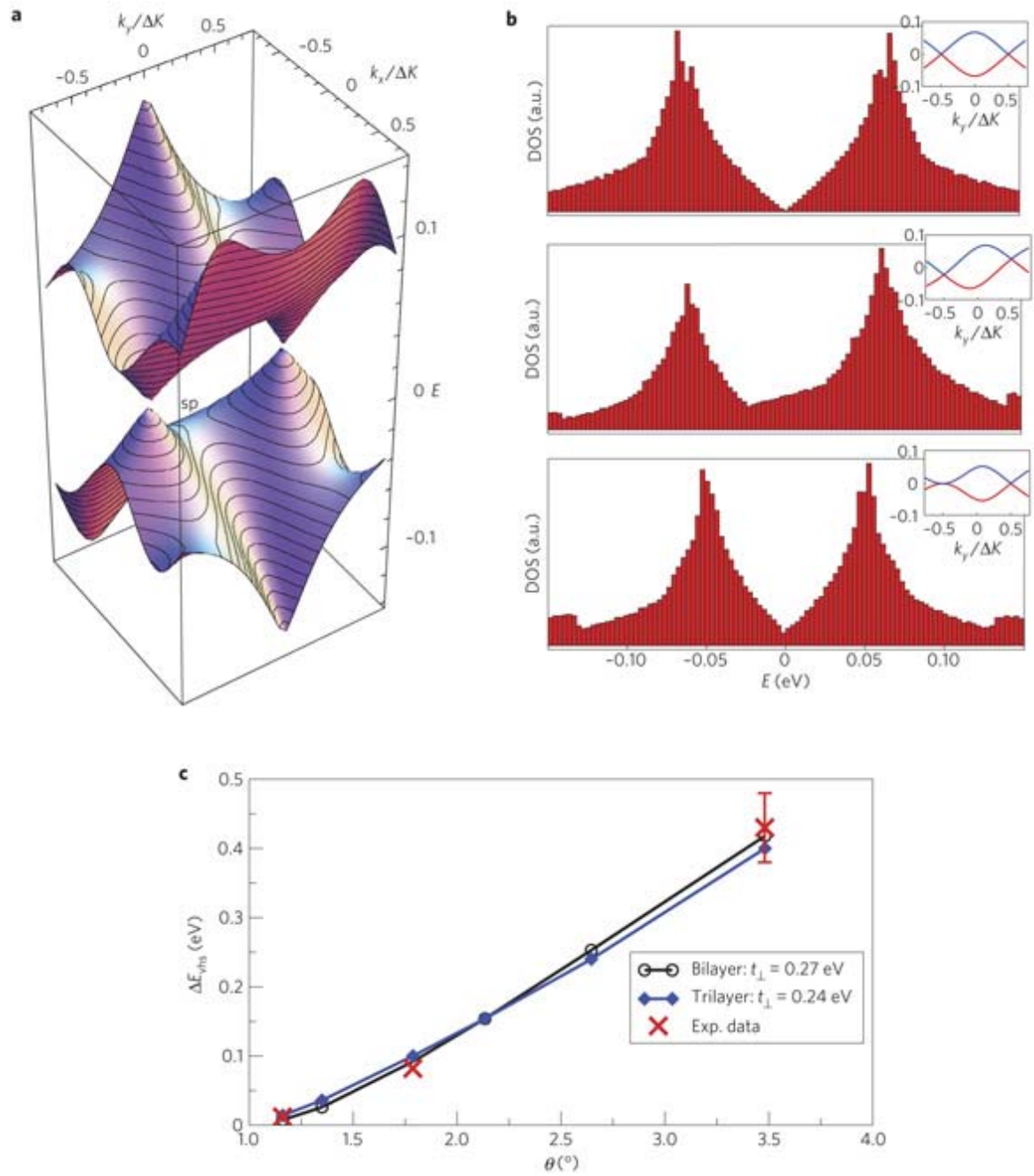

Supplementary Information for

# Observation of Van Hove singularities in twisted graphene layers

Guohong Li[1], A. Luican[1], J.M. B. Lopes dos Santos[2], A. H. Castro Neto[3], A. Reina[4], J. Kong[5]

and E.Y. Andrei[1]

[1] *Department of Physics and Astronomy, Rutgers University, Piscataway, NJ 08855*

[2] *CFP and Departamento de Física, Faculdade de Ciências Universidade do Porto, 4169-007 Porto, Portugal*

[3] *Department of Physics, Boston University, 590 Commonwealth Avenue, Boston, MA 02215*

[4] *Department of Materials Science and Engineering, MIT, Cambridge, MA 02139*

[5] *Department of Electrical Engineering and Computer Science, MIT, Cambridge, MA 02139*

## 1. Experimental details

Experiments were carried out in a home-built low-temperature (4K/2K) high-magnetic-field (13T/15T) scanning tunnelling microscope developed in our group. The controller was a commercial SPM 1000 unit from RHK Technologies. The tunnelling tips were obtained from mechanically cut Pt-Ir wire (from nanoScience Instruments). The magnetic field was applied perpendicular to the sample surface with a superconducting magnet working in persistent mode. The d$I$/d$V$ measurements were carried out with a lock-in technique using a 340 Hz a.c. modulation of the bias voltage. All data shown in main text were taken at 4.4K. Large area coarse nanopositioners of STM were used to search for the moiré pattern.

HOPG samples (grade 2 from SPI supplies) were cleaved in air shortly before experiments. CVD graphene layers were grown on Ni according to ref.[13]. The CVD graphene sample was transferred to Au grid and then cleaned in $H_2$ at 250°C.

## 2. Tight binding model for twisted graphene bilayers

The simplest tight-binding model of single-layer graphene considers hopping amplitudes between nearest neighbor atoms, $t \approx 3eV$. To model the bilayer, one adds the hopping amplitude between A1 and B2 atoms[6], $t_\perp \sim t/10$. The conduction and valence bands still touch at the Dirac points, but the dispersion becomes quadratic, due to the hybridization of the linear bands of each layer, which are degenerate. In the twisted bilayer, we maintained hopping amplitudes from atoms of one layer to the closest atoms (of either sublattice) in the other layer. The corresponding positions can be worked out from the geometry, and the hopping amplitudes were calculated using a parameterization of Slater-Koster integral for graphite[S1]. One arrives at a position dependent hopping $t_\perp^{\alpha\beta}$ (r) (α = A2, B2, β = A1, B1) with the periodicity of the Moiré superlattice.

When there is a rotation, the Dirac points of the two layers no longer coincide: there is a shift in the Dirac wave vector of the rotated layer to $K^\theta = K + \Delta K$. In the continuum description of a single layer, momentum is measured relative to the Dirac value K; if we use the same origin of momentum for the two layers, such that a uniform hopping couples states of the same momentum in both layers (Fig.S1), the zero energy states do not occur at k=0, but rather at k=-ΔK/2 in layer 1 and k=ΔK/2 in layer 2 (ref.11).

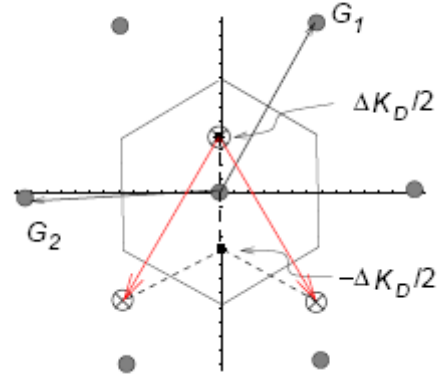

FIG.S1. Geometry of the Brillouin zone of the Moiré superlattice. G1 and G2 are the basis vectors of the reciprocal lattice of the Moiré super lattice.

The position dependent hopping can be analyzed by its Fourier amplitude $t_\perp^{\alpha\beta}$ (G), which couple states of momentum k and k+G. These Fourier amplitudes were calculated in reference (11); for small angles, three values of G, (G=0, G= -$G_1$, and G=-$G_1$-$G_2$ where $G_1$ and $G_2$ are reciprocal superlattice vectors) are found to give the dominant amplitudes; they have the same modulus, which, for small angles is practically constant, $t^\theta_\perp \approx 0.4\, t_\perp$.

The VHS in Fig. 4a can be understood as follows. The low energy states of, say, layer 2 occur near k = $\Delta$K/2. They are coupled by the inter-layer hopping to states of layer 1 at three momentum values, represented by the symbol $\otimes$ in Fig.S1: k =$\Delta$K/2, $\Delta$K/2-$G_1$, $\Delta$K/2-$G_1$-$G_2$; these three momentum values are at the same distance, $\Delta$K=K×2sin($\theta_i$/2) from the Dirac point of layer 1; the corresponding energies are ±$v_F\Delta$K. We can therefore perform a simple calculation of the lowest lying bands, in the manner of a quasi-free electron approximation, by including plane waves k, k+$G_1$ and k+$G_1$+$G_2$ in layer 1 and k, k−$G_1$ and k−$G_1$ −$G_2$ in layer 2 (twelve states overall, including positive and negative energies).

The momentum shift between the two Dirac cones turns out to be the essential difference between this problem and that of the unrotated bilayer. In the latter, the degeneracy points of both layers occur at the same momentum and the inter-layer hopping couples two doublets of zero energy states. In the present case, we have one doublet of zero energy states coupling to three pairs of states at finite energies, ±$v_F\Delta$K. As a result, the linear dispersion near zero energy is retained. However, near the center of the superlattice Brillouin zone the cones intersect. The resulting energy degeneracy is lifted by the couplings between the layers $t_\perp^{\alpha\beta}$ (G). A saddle point is unavoidable as the energy has a maximum along the line joining K and $K^\theta$ and a minimum along the intersection of the two cones. The Density of States has been calculated numerically by integration over the superlattice Brillouin Zone. To model the trilayer, we added a third layer with the same orientation as one of the previous ones, to which it couples with a zero G coupling $t_\perp^{AB}$= $t_\perp$.

## 3. Scanning tunneling spectroscopy of Moiré pattern in magnetic fields

Contrary to the naïve expectation that a small twist between layers would decouple them to reveal the electronic structure of an isolated graphene layer; we find that it drastically changes the density of states (DOS). In zero field the DOS develops sharp Van Hove singularities (VHS). In finite field the Landau level sequence of the twisted graphene at small rotation angles no longer follows the square root dependence in field and level index expected of the of massless Dirac fermions in single layer graphene. In the case of the decoupled layer the Landau levels consist of a single sequence with square root dependence on field and level index[16] in stark contrast to the twisted layer spectra where the Landau level sequence is seen as a small modulation on top of the VHS peaks. At large angles, the square-root sequence is seen in between the VHS but with a reduced Fermi velocity. (to be reported elsewhere).

In a recent report[17], the observation of a Moiré pattern and a sequence of Landau levels on epitaxial graphene on SiC(000-1), was attributed to a twist-induced decoupling of the graphene layers. However, in view of the results presented here and in reference 16, this interpretation is incorrect. Although the period of the Moiré pattern in some regions of reference 17 was comparable to that observed here, no VHS were seen. More surprisingly the Landau levels have been shown[17] independent of the period of the moiré pattern. In fact, the Landau levels should depend on the twist angles for the moiré patterns shown in reference17 and the dependence becomes weak only for very large angle rotation (to be reported elsewhere).

## 4. Large angle moiré pattern in CVD graphene

Fig. S2 shows a moiré pattern in CVD graphene sample with a rotation angle of ~ 3.48°, corresponding to i=9. Scanning tunneling spectrum also displays two VHS but with a larger separation of ~430mV. In contrast to the small angle data, there is no detectable difference in tunneling spectrum at peak or valley of moiré pattern, Fig.S3 (next page), indicating absence of charge density wave at large twisting angle.

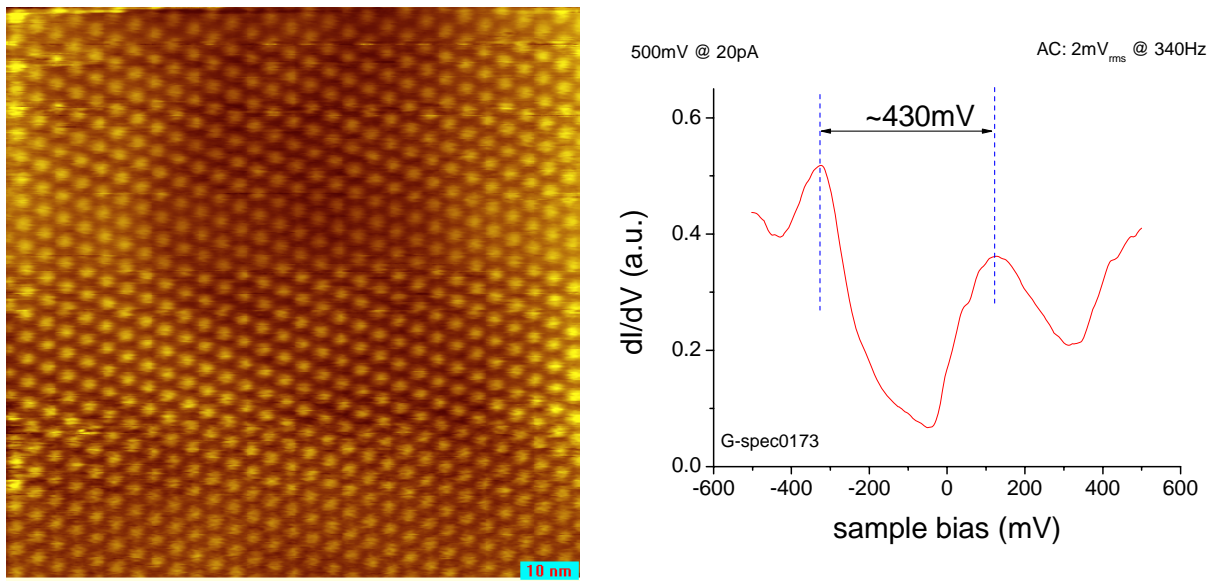

**FIG.S2.** Moire pattern in CVD graphene sample. (left) Topography of a moiré pattern with lattice constant of 3.9±0.2 nm. Tunneling current 20pA, sample bias voltage 200mV. (right) tunneling spectrum showing two VHS separated by ~430mV.

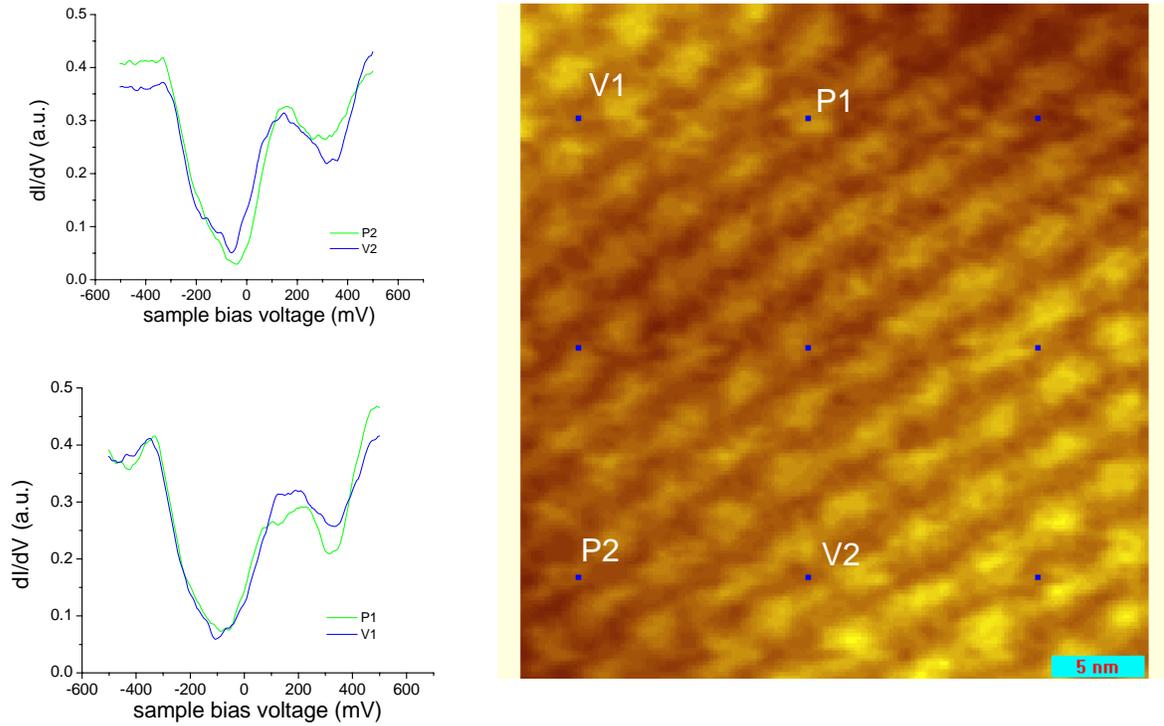

**FIG.S3.** Scanning tunneling spectra (left two panels) taken at different positions in Moire pattern (right) are identical within experimental noise, showing the absence of localization near VHS at large twisting angle.

## References

S1. E. V. Castro et al arXiv:0807.3348.